\documentclass[useAMS,usenatbib]{mn2e}
\usepackage{graphicx}
\title[The EOS of dense matter from pulsars observations]
{Constraints on the dense matter EOS from the measurements 
of PSR J0737-3039A moment of inertia and PSR J0751+1807 mass}

\author[ M.Bejger, T.Bulik, and P. Haensel]{
M. Bejger\thanks{E-mail:bejger@camk.edu.pl},
T. Bulik\thanks{E-mail:bulik@camk.edu.pl}
and P.Haensel\thanks{E-mail:haensel@camk.edu.pl}\\
Nicolaus Copernicus Astronomical Center, Bartycka 18, 00716 Warsaw, Poland  }

\begin{document}
\bibliographystyle{mn2e}
\date{Accepted .... Received ....; in original form .....}

\pagerange{\pageref{firstpage}--\pageref{lastpage}} \pubyear{2005}

\maketitle

\label{firstpage}

\begin{abstract}
The moment of inertia of the pulsar A in the neutron star binary 
J0737-3039 will soon be measurable through detailed
measurements of the periastron advance. We present the
calculation of the moment of inertia  of neutron   stars with
the masses of the components of the binary J0737-3039
for a broad range of equations of state of dense matter and
discuss the implications of such measurement for constraining
the equation of state. An observational determination of the
moment of inertia of the pulsar A in J0737-3039 with the
accuracy of $10\%$ shall narrow down considerably the range of
viable equations of state. We also show that limits on maximal
mass of a neutron star provide a complementary set of
constraints on the properties of dense nuclear matter.
\end{abstract}


\section{Introduction}

Constraining the equation of state (EOS) of dense matter in
the interior of neutron stars is a fundamental problem of
modern astrophysics. At present, many different theoretical
EOSs at densities exceeding nuclear density ($\rho_0~=~2.5~\times~
10^{14}~{\rm g~cm^{-3}}$) are available. Their predictions
diverge at a few times $\rho_0$, which reflects poor knowledge
of the EOS of neutron star cores. To constrain the EOS, one
needs to measure neutron star masses and radii or some
functions of these quantities. This is an extremely difficult
task.

Neutron star masses are best known for the radio
pulsars residing in binaries. The estimates of their masses
 have been
achieved by measuring the post-Keplerian parameters of binary
motion. The best studied cases are the double neutron star
binaries, like e.g. the Hulse-Taylor pulsar, or the recently
discovered binary pulsars J0737-3039A,B
\citep{2003Natur.426..531B}. Masses have also been strongly
constrained in some binaries with white dwarfs
\citep{Nice:2004fn,Nice2004IAUS}. Masses of the accreting neutron
stars have been constrained by measuring the motion and
spectral properties of the companions. The latter, however,
suffer from larger systematic uncertainties because of
difficulty in modeling atmospheres of stars illuminated by X
and gamma rays from the neutron star.

Constraining a second parameter like e.g. radius or any function of the
mass and radius is much more difficult. One can achieve that
by measuring the thermal UV or soft X-ray radiation from the surface
of a neutron star \citep{2003tsra.symp..319P}.
 The estimates of the radius depends in this case
on precise modeling of the spectrum in the presence of strong
magnetic field and with unknown chemical composition of the
atmosphere. The constraints on the radius depend also on the
estimate of the distance to the source. Identification of
spectral lines in the soft X-ray spectra would provide an
estimate of the gravitational redshift of a neutron star.
However, no such line have been reliably identified yet
(results of \citealt{2002Natur.420...51C} still need a confirmation).
The discovery of kHz QPOs raised hopes to constrain masses and
radii of neutron stars. These oscillations are most likely
connected with motion in the inner disk around a neutron star.
The estimates based on the QPO frequencies depend on a
particular model of the QPO phenomena and also are uncertain
due to erratic behavior  of the oscillations.

The discovery of the binary pulsars J0737-3039A,B provided an opportunity
 to test General Relativity and pulsar physics in a unprecedented detail.
 The system is highly relativistic, and only a few months of observations
led to the constraints similar to those of the Hulse-Taylor
system after a decade \citep{2004Sci...303.1153L}. It is a
nearly edge on binary which provides an opportunity to
investigate the region inside the light cone of pulsar B. The
measurements of the pulsar motion are precise enough to
measure the second order post-Newtonian (2PN)  effects. At
this level of accuracy the periastron advance depends also on
the spin-orbit coupling which involves the moment of inertia
of pulsar A. The possibility of using this effect to measure a
neutron star moment of inertia was first noted by
\citet{1988NuoCi..101.127D}. \citet{2004astro.ph.11470L} 
estimated that the moment of inertia of pulsar A should be
measurable with the accuracy of about $10\%$. 
\citet{2004ApJ...617L.135M} presented a
calculation of the moment of inertia for three classes of
equations of state and showed that such measurement would
allow to distinguish between these classes.

The moment of inertia of a neutron star is a very
interesting quantity as it involves the square of the radius.
 Therefore, for a given mass,  it  is very sensitive
 to the details of the EOS, and in particular its stiffness.
 Sensitivity of $I(M)$ to the EOS was already utilized
in the analyses of the energetics of the Crab Nebula
expansion, powered by the Crab Pulsar at the expense of the
neutron star  rotation energy. Resulting values of $I$ were
rather high, indicating a stiff EOS \citep{BH2002,BH2003}.
This result is in accordance with very recent
measurements of the mass of the PSR J0751+1807, which at the
95\% confidence level is $2.1^{+0.4}_{-0.5}~{\rm M}_\odot$
 \citep{Nice:2004fn}.

In the case of PSR J0737-3039 A the
gravitational mass is already very precisely known
($1.338~{\rm M}_\odot$) and perspectives of the future
measuring of $I_{\rm A}$ via analysis of the effect of the
spin-orbit coupling in the pulsar timing motivate
a detailed calculation of the moment of inertia.
 Here we use a broad set of EOSs  in order to
study various correlations involving the type of the EOS and
$I_{\rm A}$, and discussing observational constrains from measurements
of  stellar moment of inertia  and mass.
 In section 2 we briefly describe the
EOSs we use, and we present the method of calculating
the moment of inertia. Section 3 contains the results,
while in section 4 we present the summary and discussion.

\section{Calculating moments of inertia}

\subsection{Description of EOSs}

To illustrate the broad range of possible values of the moment
of inertia, we use a set of twenty three EOSs of dense matter
available in the literature.   Similarly as
\citet{2004ApJ...617L.135M}, we divide the EOSs into three
classes. The first class contains EOSs involving only nucleons
and leptons, based on realistic nuclear interactions fitted to
nuclear-physics data, and obtained using modern, precise
solutions of the many-body problem ({\bf Class I} : APR
 of \citealt{APR}, WFF1-3 of \citealt{Wiringa1988}, BBB1-2 of 
\citealt{Baldo1997}, SLy of \citealt{SLy}, FPS of \citealt{FPS}). 
Out of these EOSs, only FPS and APR models were considered 
by Morrison et al. (2005). We added  two EOSs based on the 
Brueckner-Bethe-Goldstone theory of nuclear matter (BBB1, BBB2), 
one EOS derived from the Skyrme effective nucleon-nucleon interaction, 
consistent with variational calculations for dense neutron matter 
(SLy), and three EOSs (WFF1-3) obtained in variational calculations of Wiringa
et al. (1988).   

 A second class contains EOS with less solid nuclear and
many-body basis, but allowing for more speculative
extrapolations : non-relativistic effective forces and
relativistic mean-field models involving nucleons and
hyperons, and models with phase transitions to kaon
condensates and quark matter ({\bf Class II}: GNH1-4
 of \citealt{Glend1985},  WGW1-2C of \citealt{Weber1991}, BGN1H2, BGN2H1,
BGN2H2 of \citealt{Balberg1997,Balberg1999}, GNHQ of \citealt{Glend_book1997} 
(mixed-phase hadron-quark EOS, model 2) and PREPL of \citealt{Pons2000}).
This set of EOSs is much larger and richer than Class II of Morrison 
et al. (2005), who restricted themselves to two EOSs obtained using
the Relativistic Mean Field model. 

 Finally, third class is the most
exotic, as it includes models of self-bound strange (quark)
stars ({\bf Class III}: DBDRS of \citealt{Dey1998} and SM, SM60
 of \citealt{Zdunik2000}). This class contains only three EOSs
because the scaling relation (see Sect.3) enables one to easily predict
results for any EOS of strange quark matter intermediate
between the stiffest SM60 EOS and the softest DBDRS EOS.

Moreover, we considered additionally three  specific EOSs of 
extremal character. Two of them are stiff
(MFT17, \citealt{MFT17}) and  soft (BPAL12, \citealt{BPAL12})
extremes of the nucleonic EOSs; they set two boundaries for
the baryonic EOSs. While they reproduce experimental
saturation density, binding energy, and symmetry energy of
nuclear matter, they strongly overestimate (MFT17) or
underestimate (BPAL12) nuclear matter incompressibility at
saturation. We also consider the CLES EOS that does not
belong to any of above classes I-III. The causal-limit (CLES)
EOS maximizes the maximum allowable stellar mass. This EOS
assumes the SLy  EOS below density
$2\rho_0=5\times 10^{14}~{\rm g~cm^{-3}}$ and the stiffest
possible (sound speed equal the speed of light) EOS at higher
 density.
%%%%%%%%%%%%%%%%%%%%%
\subsection{Numerical technique}
%%%%%%%%%%%%%%%%%
 The moment of inertia of a neutron star
  is calculated assuming  rigid rotation, and
 in general case it is defined as $J/\Omega$, where $J$ is the
 stellar angular momentum and
$\Omega$ is the angular frequency of rigid rotation as
measured by a distant observer. If $\Omega$ is much smaller
than the mass-shedding (Keplerian) limit for a given mass,
$\Omega_{\rm ms}$, then a slow rotation approximation is
valid, and  $I$  can be calculated using non-rotating stellar
model \citep{Hartle1967}. In the case of J0737-3039 A pulsar
$\Omega=276.8 {\rm s}^{-1}$ is less than one tenth of
$\Omega_{\rm ms}$, and therefore the Hartle approximation
should be valid within a percent (corrections to it are
$\propto (\Omega/\Omega_{\rm ms})^2$).

\begin{figure*}
\begin{center}
  \includegraphics[width=1.90\columnwidth,angle=0]{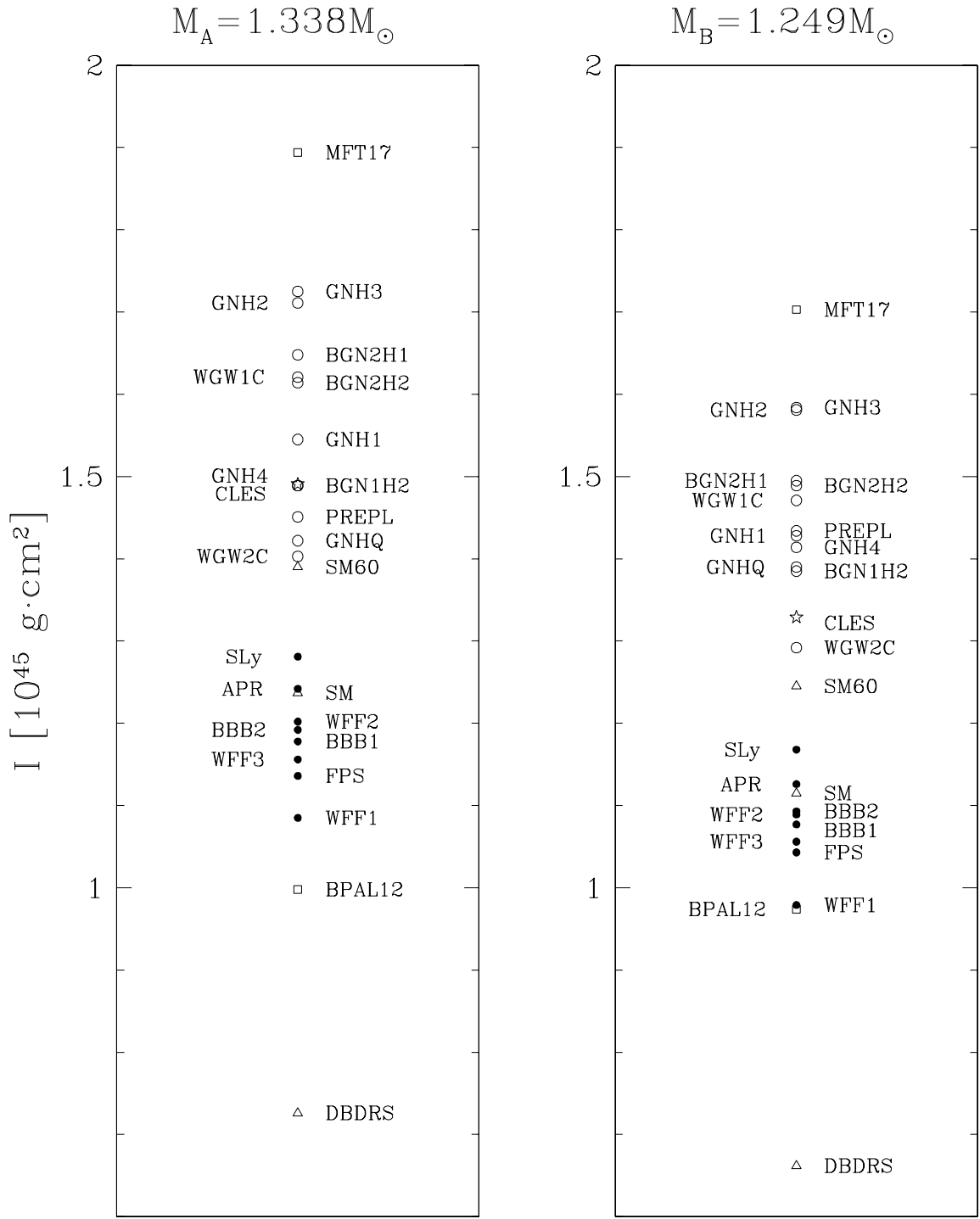}
 \end{center} \caption{The moments of inertia of the binary pulsar
  components: A- left panel, and B - right panel.
  Filled circles: Class I (realistic EOSs involving nucleons and leptons).
  Open circles: Class II (relativistic and non-relativistic
  mean-field models involving hyperons,   kaon condensates and
  quark matter). Open triangles: Class III  - strange quark stars;
  the range of  values of the moment of inertia available for strange stars
  lies between the triangles labeled DBDRS and SM60.
  Open squares: extreme nucleonic EOS. Open star: causal-limit EOS
  (CLES).
 For further
  explanations see the text. }
  \label{moi}
\end{figure*}

The values of moments of inertia for hydrostatic stellar models
based on the presented EOSs were obtained while solving the
Tolman-Oppenheimer-Volkoff equation \citep{Tolman39,OppVolk39}
 for spherically symmetric star,
 within the slow-rotation approximation \citep{Hartle1967}.
Additionally, for a representative
 subset of EOS we have directly checked the accuracy of this
 approximation by
 computing the moment of inertia for the axisymmetric rotating
stellar model within the highly accurate 2-D LORENE framework
\citep{BGSM93,BGM1998}. The virial error indicators which
describe the accuracy of computation were typically $10^{-5}$
or less. As one could suspect, the difference between
non-rotating and axisymmetric rotating models at this rotation
rate is of the order
 of $\sim 1\%$ or less. It is principally dominated by the difference
 in methods of interpolating
the tabulated EOS. Therefore, we draw the conclusion that the
 spherically symmetric
 calculations are perfectly suitable for reasonable
 estimation of the moment of inertia
 for 
J0737-3039 
components.
 The results for the slow-rotation approximation
 for both pulsar components of the  J0737-3039 system are displayed
 in Fig.~\ref{moi}, and the precise parameters for a small subset of EOSs
 (for pulsar A) are collected in Table~\ref{tabval}.

%------------------------------------------------
\begin{table}
 \centering
 %\begin{minipage}{140mm}
  \caption{Highly accurate 2-D results for rotating axisymmetric stellar models of
 gravitational mass $M=1.338~M_\odot$ and rotational frequency
 $\Omega=276.79~{\rm s^{-1}}$ computed using the LORENE subroutines for a
 small subset of EOSs.
The symbol $\rho_{\rm c,14}$ denotes
 the central density (in units of $10^{14}~{\rm g\,cm^{-3}}$), 
$J_{47}~[10^{47}~{\rm cm^2\,g\,s^{-1}}]$ - the angular momentum, 
$I_{45}=J/\Omega~[10^{45}~{\rm g\,cm^2}]$ the
moment of inertia and $R_{\rm circ}$ the circumferential radius (in km). 
In general,
the values of the moment of inertia differ by approximately 1\% from the
non-rotating results - the difference is dominated by the error of
the tabulated EOS interpolation method.}
  \begin{tabular}{ccccc}
  EOS & $\rho_{\rm c,14}$
      & $J_{47}$
      & $I_{45}$  
      & $R_{\rm circ}$ \\ \hline\hline
  BPAL12 & 19.57 & 2.72 & 0.98 & 10.52 \\ \hline
  APR & 9.74 & 3.39 & 1.23 & 11.35 \\ \hline
  SLy & 9.39 & 3.57 & 1.29 & 11.76 \\ \hline
  BGN2H1 & 6.09 & 4.53 & 1.64 & 13.47 \\ \hline
  GNH3 & 6.18 & 4.77 & 1.72 & 14.28
\label{tabval}
\end{tabular}
%\end{minipage}
\end{table}

\section{Results}

%%%%%%%%%%%%%%%%%%%%%%%%%%% PH %%%%%%%%%%%%%%%%%%%%%%%%%%%
 The dependence of the
predicted values of  $I_{\rm A}$ and $I_{\rm B}$ on the
assumed EOS of dense matter is visualized in Fig. 1. Within
the considered set of EOSs $I_{\rm
A,45}=0.7-1.9$ and $I_{\rm B,45}=0.7-1.7$, a huge scatter indeed 
(we adopt the notation $I_{\rm 45}\equiv I/{\rm 10^{45}~g~cm^2}$). 
However, if we restrict
ourselves to most recent EOSs consistent with experimental
nuclear matter parameters and involving only nucleons, then
the range of predicted values narrows considerably, to $I_{\rm
A,45}=1.1-1.3$ and $I_{\rm B,45}=1.0-1.2$. The EOSs which
involve hyperons and/or exotic phases of dense matter (quark
and kaon condensate) at supra-nuclear densities turn out to be
so stiff for neutron stars with the masses of the pulsars  A
and B  that they yield $I_{\rm A,45}=1.4-1.7$  and $I_{\rm
B,45}=1.3-1.6$. 
Their high stiffness  within  
PSR J0737-3039A,B results from the requirement that they 
should produce maximum allowable mass $>1.44~{\rm M}_\odot$. 
The maximum mass is sensitive to the 
 high-density segment of the EOS, with strong softening due to hyperons
or a phase transition. For this segment to be sufficiently stiff to 
respect consistency with observations, the lower-density segment 
(without hyperons and/or new dense phase)  should be indeed very stiff.
 The two extremes of nucleon  EOSs set are
MFT17 (very stiff) and BPAL12 (very soft); while they
reproduce experimental saturation density, binding energy, and
symmetry energy of nuclear matter, they strongly overestimate
(MFT17) or underestimate (BPAL12) nuclear matter
incompressibility. 

The values obtained for the causal-limit CLES EOS (open star symbols) 
are similar to those obtained for the Class II EOSs. At relatively 
low neutron star masses, the SLy envelope with $\rho<2\rho_0$ is  
important for stars structure. However,  the core with $v_{\rm sound}=c$ 
shifts up the values of $I_{\rm A,B}$,  by more than twenty percent, 
compared to the SLy ones.

A separate subset of EOSs (open triangles)
is composed of three models of self-bound strange quark
matter. The values of $I_{\rm A,45}$ and $I_{\rm B,45}$
predicted by the strange-star models are very scattered
$0.7-1.7$. However,  this range of values can be easily
understood in terms of the scaling properties of $I$ and $M$
with the value of the surface density of strange stars.
 To a very  good approximation the values of $I_{\rm A}$ and $I_{\rm B}$
calculated for different  EOSs of strange quark matter are
related by a simple relation  involving surface  density of
bare strange stars (= density  at zero pressure), $\rho_{\rm
s}$. Namely, $I_{\rm A}$ calculated for the EOS with zero
pressure density $\rho_{\rm s}$ is related to $I_{\rm
A}^\prime$, calculated for the EOS with $\rho_{\rm s}^\prime$,
by $I_{\rm A}^\prime\simeq (\rho_{\rm s}/\rho_{\rm
s}^\prime)^{2/3} I_{\rm A}$ (this relation is obtained using
constant-density Newtonian star models).  Therefore, huge
disparity in $I_{\rm A}$ for the DBDRS and SM60 EOS can be
accounted for by the difference in $\rho_{\rm s}$. We
considered bare strange-star
models of J0737-3039 A, B. However, the mass of the crust of
strange stars is at most $10^{-4}~M_\odot$, and therefore its
contribution to $I$ is at most  0.01\% and can be safely
neglected.
%%%%%%%%%%%%%%%%%%%%%%%%%%%%%%%%%%%%
\begin{figure}
\includegraphics[width=0.95\columnwidth]{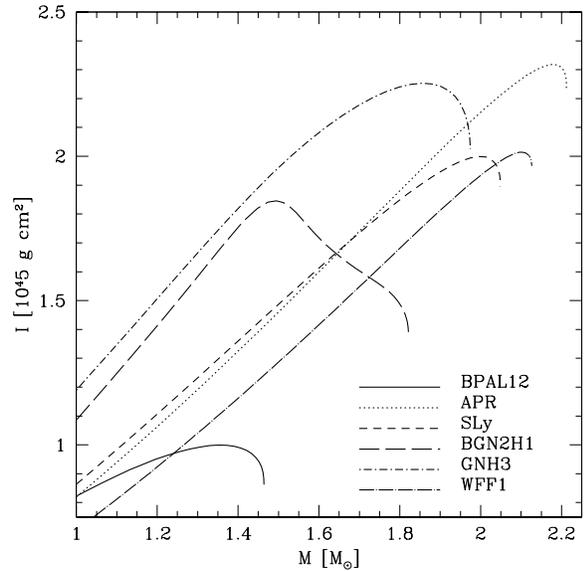}
\caption{The dependence of the moment of inertia on the
gravitational  mass for a few representative equations of
state.} \label{IofM}
\end{figure}

\begin{figure*}
\begin{center}
\includegraphics[width=1.9\columnwidth,angle=0]{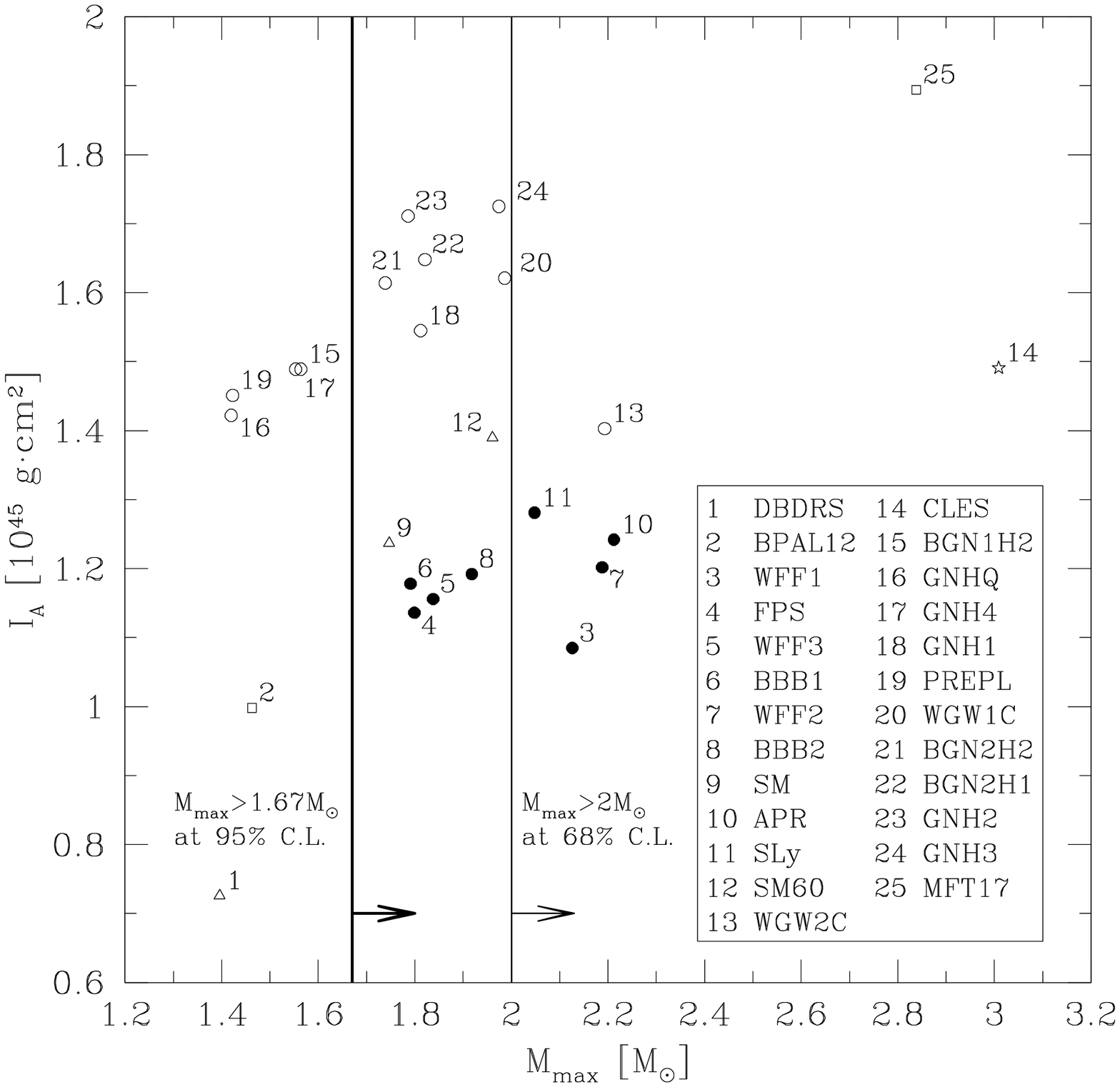}
\caption{Moment of inertia of a neutron star with mass of pulsar A versus
the maximum mass of a neutron star plotted for the same set of EOSs
as in Fig.~1. The vertical thin solid line denote the 
 68\% confidence lower limit on the maximum mass ($2.0\;$M$_\odot$), 
whereas the thick solid line marks the 95\% confidence 
 limit ($1.67\;$M$_\odot$) - the limits were 
 obtained from the PSR J0751+1807 
 mass measurements (see text for details).}
\label{IvsMmax}
\end{center}
\end{figure*}

\section{Summary and discussion}

We have presented the moments of inertia of neutron stars
J0737-3039 A, B for 25 different EOSs of dense matter.
Inspection of Fig. 1 leads to conclusion that a measurement
 of $I_{\rm A}$ with a 10\% precision would necessarily imply rather
precise picking up of the EOS of matter in  J0737-3039 A. It
could be a stringent test of ``realistic EOSs'' of Class I.
Either the EOS inside  J0737-3039 A is stiffer than our best
nuclear models predict and belongs to Class II, or is in
harmony with nuclear many-body theory represented at its best
by Class I. We therefore confirm and extend  
the result of \citet{2004ApJ...617L.135M}.  
While it would be difficult to eliminate strange star
model altogether, measurement of $I_{\rm A}$ would strongly
limit the value of $\rho_{\rm s}$ of strange quark matter at
zero pressure. The measurement of the moment of inertia
$I_{\rm A}$ will be sensitive to the properties of the EOS for
the densities typically in the range of the central density of
the pulsar A. From Table~1 we see that this corresponds $\rho
< (6-10)\times 10^{14}$g\,cm$^{-3}$, i.e. or $\rho <
(2-4)\times \rho_0$ where $\rho_0$ is the nuclear density (see
Sect. 1). In this estimate of the density we have neglected
the unrealistically soft BPAL12 model.

It should be stressed that hopefully stringent constraints on
$I_{\rm A}$ do not probe  the EOS at densities crucial for the
value of maximum allowable mass for neutron stars, $M_{\rm
max}$. A good example of two EOSs which yield very similar
values for $I_{\rm A}$ and $I_{\rm B}$ but very different
stars at $M_{\rm max}$ are BPAL12  and WFF1 (Fig. 2). However,
as Fig. 3 shows, existence of $2.1~{\rm M}_\odot<M_{\rm max}$
neutron star \citep{Nice:2004fn} combined with determination,
within 10\%,  of $I_{\rm A}$, would quite precisely establish
the complete EOS of dense matter inside neutron stars. Lower
limits on the maximum mass of a neutron star are sensitive to
the properties of the EOS at the density $\rho \approx
(6-10)\times \rho_0$, much larger than densities relevant for
the constraints from measuring $I_{\rm A}$. Both constraints
together  would constitute an ultimate test of the many-body
theories of dense matter.

While the measurement of the moment of inertia of the pulsar A
in the J0737 system is approaching it is interesting to
speculate about further investigations of the neutron star
equation of state. Measuring the moment of inertia of the
pulsar B in this system does not look feasible with the
current instruments. However precise monitoring with the next
generation instruments like e.g. the Square Kilometer Array
may yield highly accurate data that enable determination of
$I_{\rm B}$, or moments of inertia of neutron stars in other
binary pulsar systems. The Parkes Multibeam Survey may also
bring discoveries of other highly relativistic binary pulsars,
and provide measurements of the moment of inertia of neutron
stars with other masses. One should note recent measurements
of masses in neutron star white dwarf binaries which hint at
the existence of neutron stars with masses as large as
$2.1\,M_\odot$ \citep{Nice:2004fn,Nice2004IAUS}. Such measurements
provide lower limits on the maximum mass of a neutron star and
thus are complementary to the measurements of the moment of
inertia. Limits on the neutron star maximum mass probe the
densest segment of the  EOS. We illustrate the strength of the
constraints coming from the measurement of the moment of
inertia of pulsar A in J0737-3039 system together with limits
on the maximum neutron star mass in Fig.~\ref{IvsMmax}. The
two values - $I_{\rm A}$ and $M_{\rm max}$ are not correlated
since they probe different ranges of densities of the EOS. A
simultaneous measurement of the moment of inertia of pulsar A
with accuracy of $\approx 10\%$ with a limit on the maximum
mass of neutron star approaching $2.0\,M_\odot$ should narrow
down the possible set of nuclear matter EOS to just a few
possibilities in the near future.
We also plot the lower limits on the maximum 
mass of a neutron star implied by the recent measurement of the 
mass of the pulsar PSR J0751+1807 \citep{Nice:2004fn}.
We have obtained the limits shown in the Figure by assuming 
that the errors on the mass are approximately gaussian. 
We have then integrated the approximate probability 
and obtained the 68\% lower limit on the maximum mass of 
$2.0\;$M$_\odot$, and the 95\% limit of $1.67\;$M$_\odot$.
These values are plotted as vertical lines in Fig.~\ref{IvsMmax}.
 Thus a 
simultaneous measurement of the moment of inertia of pulsar A
with accuracy of $\approx 10\%$ with a limit on the maximum
mass of neutron star approaching $2.0\,M_\odot$ shall narrow
down the possible set of nuclear matter EOS to just a few
possibilities in the near future.

\section*{Acknowledgments}

This work was supported by the following KBN grants
2\,P03D\,001\,25 (TB) and  1\,P03D\,008\,27 (PH and MB).

\newcommand{\apj}{ApJ}
\newcommand{\apjs}{ApJS}
\newcommand{\apjl}{ApJ Lett}
\newcommand{\nat}{Nature}
\newcommand{\aap}{A\&A}
\newcommand{\prc}{Phys.~Rev.~C}
\newcommand{\prd}{Phys.~Rev.~D}
\newcommand{\physrev}{Phys. Rev.}

\bibliography{inertia}

\label{lastpage}

\end{document}